\documentclass[twocolumn,final,twoside,journal]{IEEEtran}

\usepackage[usenames,dvipsnames]{xcolor}
\usepackage{amsmath,amsthm,graphicx,cite}
\usepackage{amssymb}
\usepackage{epsfig,amsfonts}
\usepackage{graphicx,cite,amssymb,amsmath}
\usepackage{color}
\usepackage{mathtools}

\usepackage{enumitem}
\usepackage{dblfloatfix}
\usepackage[nolist]{acronym}
\usepackage{psfrag}
\usepackage{perso}
\usepackage{xcolor}
\usepackage{blindtext}
\usepackage[keeplastbox]{flushend}
\usepackage{relsize}
\usepackage{mathtools}
\mathtoolsset{showonlyrefs}
\usepackage{placeins}

\newcommand{\library}{\mathcal{F}}
\newcommand{\demands}{\mathcal{D}}
\newcommand{\libraryf}[1]{{f}_{#1}}
\newcommand{\encodedf}[1]{e_{#1}}

\newcommand{\load}{\mathsf{L}}
\newcommand{\lmds}{L_1^{\text{MDS}}}
\newcommand{\lmdsdue}{L_2^{\text{MDS}}}
\newcommand{\leccone}{L_1^{\text{ECC}}}
\newcommand{\lecctwo}{L_2^{\text{ECC}}}

\newcommand{\B}{\mathsf{B}}
\newcommand{\W}{\mathsf{W}}
\newcommand{\kzw}{b_{\mathsf{W}}}
\newcommand{\kzb}{b_{\B}}

\newcommand{\N}{\mathsf{N}}
\newcommand{\M}{\mathsf{M}}
\newcommand{\us}{\mathsf{u}}

\newcommand{\nballblack}{\mathsf{u_{\B}}}
\newcommand{\nballwhite}{\mathsf{u_{\W}}}
\newcommand{\nballred}{\mathsf{u_{\zonetwo}}}
\newcommand{\kbinblack}{k_{\B}}
\newcommand{\kbinwhite}{k_{\W}}
\newcommand{\kbinred}{k_{2}}
\newcommand{\nfrac}{{\mathsf{n_F}}}

\newcommand{\minJ}{\beta_{\mathsf{J}}}
\newcommand{\minW}{\beta_{\mathsf{W}}}
\newcommand{\minB}{\beta_{\mathsf{B}}}
\newcommand{\alfGone}{\alpha_{1}}

\newcommand{\alfB}{\alpha_{\B}}

\newcommand{\minGone}{\beta_{1}}
\newcommand{\minGtwo}{\beta_{2}}
\newcommand{\uzoneone}{\mathsf{u_{1}}}
\newcommand{\uzonetwo}{\mathsf{u_{2}}}
\newcommand{\kzoneone}{k_{1}}
\newcommand{\kzonetwo}{k_{2}}

\newcommand{\pwb}{\mathsf{P_{\mathsf{oc}}}}

\newcommand{\zoneone}{1}
\newcommand{\zonetwo}{2}

\newcommand{\Jdiff}{\ensuremath{{J}}}
\newcommand{\Ydiff}{\ensuremath{{Y}}}

\newtheorem{example}{Example}

\begin{document}

\begin{acronym}

\acro{MBS}{macro base station}
\acro{MNS}{master node station}
\acro{PMF}{probability mass function}

\acro{MDS}{maximum distance separable}
\acro{ECC}{edge coded caching}

\acro{SBS}{small base  station}
\acro{BiB}{balls into bins}
\acro{RaP}{random placement}
\acro{rv} {random variable}
\acro{MoP}{most popular placement}
\acro{S}{satellite}
\acro{R}{relay}
\end{acronym}

\title{     Coded Caching at the Edge of Satellite Networks}

\author{
   \IEEEauthorblockN{Estefan\'ia Recayte \\
    \IEEEauthorblockA{Institute of Communications and Navigation of DLR (German Aerospace Center),
 \\Wessling, Germany. Email:  \{estefania.recayte\}@dlr.de}\\
 }
}
\maketitle



\thispagestyle{empty} \pagestyle{empty}

\begin{abstract}
Caching multimedia contents at the network edge is a key solution to decongest the amount of traffic in the backhaul link. In this paper, we extend and analyze the coded caching technique \cite{Maddah:Fundamental} in an unexplored scenario, i.e. at the edge of two-tier heterogeneous networks with an arbitrary number of users. We characterize the  performance of such scheme by deriving a closed-form expression of the average backhaul load and reveal   a significant gain compared to other benchmark caching schemes proposed in the literature.
\end{abstract}



\section{Introduction}\label{sec:Intro}
Caching has emerged as one of the key technologies for next-generation wireless systems. Bringing the desired content to the edge of the network, i.e. memorizing copies of relevant information close to users, has been shown  not only to alleviate the backhaul traffic, but also to significantly reduce  latency and power consumption \cite{magcache}.

 To achieve this goal, a two-step caching strategy is typically implemented, pre-fetching the content at the edge (e.g. at  base stations, LEO satellites, relays or helpers) during network off-peak periods (\emph{placement phase}), so as to serve the users without consuming backhaul capacity  when the network is congested (\emph{delivery phase}).

Several works have recently investigated the potential benefits of caching  schemes at the edge. Interesting results were obtained in \cite{caireFemto}, where  authors compared two different placement approaches, i.e. encoded and uncoded content placement to reduce the download delay.   Caching scheme based on \ac{MDS} codes have been studied in wireless networks  to minimize the expected download time \cite{piemontese2016} or to reduce the amount of data to be sent \cite{bioglio:Globcom2015}.

Furthermore, Maddah-Ali et al. introduced in their pioneering work \cite{Maddah:Fundamental}
the concept of \emph{coded caching}, considering local caching directly on the user's device.
The idea is to  deliver coded content and leverage the user's local content to serve multiple users with a single transmission.
 This technique has spurred an extensive body of research providing a solid understanding of the potential and limitations of caching, e.g. \cite{Chao:limits, Vijith:limits}. However, coded caching has  been studied especially in setups where few cache-aided users are connected to a common server via a shared link. Instead, its potential and the trade-offs it may induce in other relevant scenarios remain unexplored to date. An example of notable practical relevance is given  by two-tier networks that foresee a satellite component, which will be an integrating part of 5G and 6G systems \cite{NTNzorzi}.
In these settings, commonly referred to as non-terrestrial networks,   terminals may not be equipped with direct satellite connectivity, and the intermediate tier is responsible for forwarding content from one end to the other.

To shed light on these relevant design aspects, the present work considers the application of
  coded caching   in a two-tier caching satellite network for an arbitrary number of users. Unlike previous work, caching is considered at the edge of the network (e.g. relays, helpers, LEO satellites) and  multiple users are connected to one or more cache-aided relays.
To analyze the system performance, we derive a closed-form expression of the  average backhaul transmission load when coded caching is in place. We compare our results with  the benchmark given by the MDS caching scheme proposed in literature and we show a reduction in backhaul transmissions.  In particular, the presented scheme  triggers a gain each time the mutual difference of   sets representing the file requested at each relay is not empty. To quantify such gain, the distribution of how users request for content is casted onto a  combinatorial \ac{BiB} setting. Due to complexity of the problem, closed form expressions  are derived when the distribution of files requested is  uniform, while we show via Monte-Carlo   that the aforementioned gain is also present when files are not equiprobable. The significant enhancement obtained encourages to investigate further relevant aspect of the edge coded caching in satellite networks.

\subsubsection*{Notation}
We use capital letters, e.g. $X$, for discrete \acp{rv} and their lower case counterparts, e.g.  $x$, for their realizations. The probability mass function (pmf) of the \ac{rv}  $X$ is denoted as $p_{X}(x)$ and  conditional pmfs as ${\Pr\{  X =  x \,|\,   Y =  y  \} = p_X( x |  y)}$. A set is denoted with calligraphic letters, e.g. $\mathcal{S}$. The cardinality of set $\mathcal{S}$ is indicated as $| \mathcal{S}|$.

\section{System Model}\label{sec:sysmodel}

\begin{figure}[t]
 \includegraphics[width=0.48\textwidth]{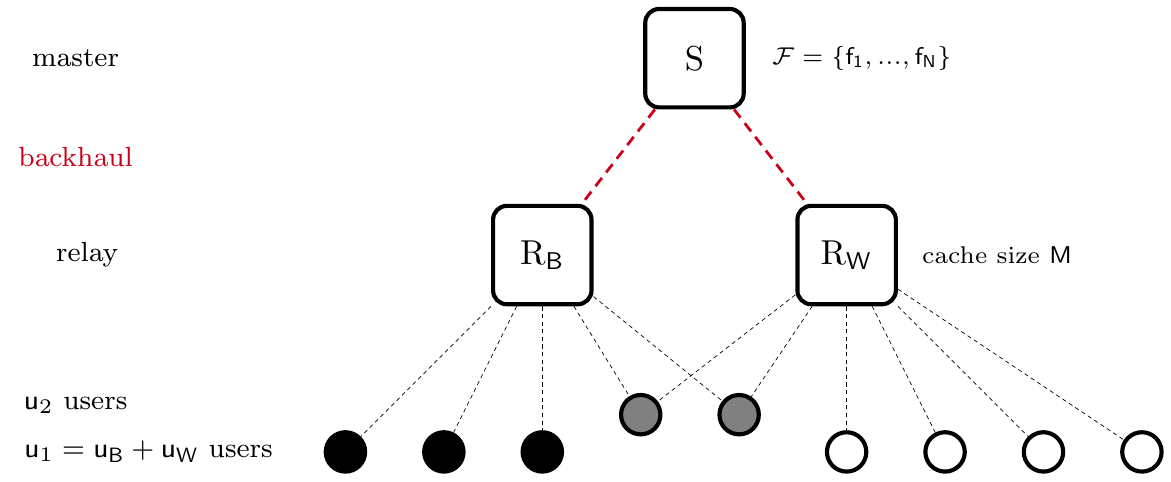}
 \centering \caption{System model: cache-aided relays are connected to the satellite through the backhaul link and users can be connected to one or more relays. }
 \label{fig:systemMo}
\end{figure}

A two-tier heterogeneous network composed  by a master node,  two cache-enabled nodes and a number of end users are considered.
While this setup applies to different network configurations, we will take as reference throughout our discussion the satellite topology illustrated in Fig.~\ref{fig:systemMo}. Here, a \ac{S} stores a whole library ${\library =\{ f_{1}, \cdots, f_{\N} \}}$ of equal-size files. On the ground, two cache-enabled  relays (R$_{\B}$ and  R$_{\W}$)\footnote{The subscript $\B$ and $\W$ have been chosen to facilitate the similarity between the caching scheme and BiB problem, as will become clear later.} are connected via a backhaul link to \ac{S}, and each one provides connectivity to some users.
 As typical in current satellite-aided terrestrial networks, we assume that no direct link between users and \ac{S} is available\footnote{Note that the setup presented is not limited to this architecture. For instance, a possible scenario may consist of a GEO satellite  which acts as master node connected via  backhaul links to cache-enabled LEO satellites.}. Depending on their locations users (terminals) may be connected to one or both relays.  $\mathcal{U}_h$   denotes the subset of users  that are connected to $h$ relays, where $h= \{1, 2 \}$, whereas $\mathcal{U}_{{\B}}$ ($\mathcal{U}_{{\W}}$) denotes the subset of users connected only to relay {R}$_{{\B}}$ ({R}$_{\W}$). Note that the set of users connected to exactly one relay is $\mathcal{U}_1 = \mathcal{U}_{{\B}} \cup \mathcal{U}_{{\W}}$.

 Each relay can store up to $M \leq \N$ files locally. With $\mathcal{Z}_{\B}$ ($\mathcal{Z}_{\W}$) we indicate the files present in the cache of   {R}$_{\B}$ ({R}$_{\W}$).
 During the \emph{placement phase}, which is carried out offline, each file $f_j \in \library$ is partitioned into $\nfrac$  equally long fragments, i.e. $ \libraryf{j}$ is fragmented as $ \{ \libraryf{j}^{(1)}, \cdots, \libraryf{j}^{(\nfrac)} \}$.  Each cache stores $F_j$ fragments related to file $\libraryf{j}$ according to one of the caching schemes that will be introduced later in this section.
During the \emph{delivery phase}, the network serves the user's requests.
Each user picks a file according to the file distribution considered (e.g. uniform  or Zipf distribution).
A user connected to $h$ relays which request for   $\libraryf{j}$ receives  ${h \, F_j}$ fragments directly from the cached content and  the $\max(0, \nfrac - h   F_j)$ missing fragments are forwarded by the R after  being retrieved by S via the backhaul link.
The transmission technique in the backhaul link depends on the caching scheme considered.  With $\demands_{_x} \subseteq \library$ we denote the subset of files requested by the set of users $\mathcal{U}_x$, where $x \in \{1,2,{\B},{\W} \}$. For example, the cardinality of  $\demands_1$ is the number of different files requested by users connected to a single relay (users in $\mathcal{U}_1$).

 In such configuration, we indicate with $\us$ the total number of terminals that concurrently request content from the library, each independently choosing a file to download. Specifically, we have that $ \us = \us_{\zoneone} + \nballred$ where $|\mathcal{U}_1| =\us_{\zoneone}$, are those  connected to a single relay,  while  $|\mathcal{U}_2| = \us_{\zonetwo}$ are those connected to both relays. We also have that  $ \us_{\zoneone} = \nballblack + \nballwhite$ where  $\nballblack$  are the users requesting content only at  {R}$_{\text{B}}$ and $\nballwhite$ only  at  {R}$_{\text{W}}$. A relay directly delivers content present in its cache and retrieves content that is not available locally via the backhaul link. For simplicity we assume that all transmissions are error-free.

Following this notation, we analyze a coded caching scheme at the edge based on the strategy proposed in \cite{Maddah:Fundamental}, referred to as the  \emph{edge coded caching scheme} (ECC). To evaluate the scheme, we derive the average backhaul transmission load $\load$, i.e. the average number of  packets that S should transmit in the backhaul link to satisfy user requests. We will focus on the rv $L^{\text x}$ which describes the number of transmissions required from  S for a given caching scheme $\text x$. We have that  $\load^{\text x} = \mathbb{E}[L^{\text x}]$ where the operator $\mathbb{E}$ indicates the expected value.
 The metric $\load$ is used to compare the behavior against the benchmark given by \ac{MDS} scheme.
 Let us discus both schemes in the following.
\vspace{-.7em}
\subsection*{ \ac{MDS} Caching Scheme}
In the \ac{MDS} scheme \cite{bioglio:Globcom2015}, the network works by caching and delivering packets that are encoded.
 In particular, $\nfrac$ fragments of file  $\libraryf{j}$  are used to create $n > \nfrac$ encoded packets using  a $(n, \nfrac)$  MDS code. The set of encoded packets related to  $\libraryf{j}$ can be  written as  ${ \encodedf{j}  =\{ \encodedf{j}^{(1)}, \cdots, \encodedf{j}^{(n)} \}}$ where $ \encodedf{j}^{(i)}$ and $\libraryf{j}^{(i)}$ are equally long for every $i$ and $j$. With the MDS coding technique, a  user can  reconstruct  successfully  the requested file by receiving any subset of $\nfrac$ encoded packets \cite{bioglio:Globcom2015}.

If we assume a uniform distribution of the file requested then it is also assumed that files are split into $\nfrac = \N$ fragments. Each relay fills own cache with $F_j = \M$ encoded packets per file such that ${\mathcal{Z}_1 \cap \mathcal{Z}_2 = \emptyset}$, i.e. relays store a different subset  of encoded packets for the same file. The satellite keeps    ${n- 2 \M}$ encoded packets for every file. The delivery phase is split into the following stages. First, users receive content from the relays' cache, subsequently the missing encoded packets are  sent  by   S through the backhaul link to the R which forwards them to the users. The benefit of this strategy is based on being able to serve both relays in parallel with a single transmission via the backhaul link. This occurs whenever there are requests for the same content at both relays. To clarify, consider the following numerical example.

\begin{example}
  Let us assume   to have two users: user $1$ is connected only to  R$_{\B}$  and   user $2$  only to {R}$_{\W}$. Consider a memory size of $\M=1$   and two equiprobable files  split into $\nfrac = 2$ fragments \[ \libraryf{1}= \{\libraryf{1}^{(1)},  \libraryf{1}^{(2)}\} \text{ and }   {\libraryf{2}= \{ \libraryf{2}^{(1)}, \libraryf{2}^{(2)}\}.} \]  Let us consider a $(3,2)$  MDS code such that  we can write the encoded packets as \[
  \encodedf{1}  =\{ \encodedf{1}^{(1)}, \encodedf{1}^{(2)},  \encodedf{1}^{(3)} \}  \text{ and  }  \encodedf{2}  =\{ \encodedf{2}^{(1)}, \encodedf{2}^{(2)}, \encodedf{2}^{(3)} \}. \]
   We further set $\mathcal{Z}_{\B} =\{\encodedf{1}^{(1)}, \encodedf{2}^{(1)}\}$ and $\mathcal{Z}_{\W} =\{\encodedf{1}^{(2)}, \encodedf{2}^{(2)}\}$.

  To characterize the average backhaul load $\mathsf{L}^{\text{MDS}}$ , we shall consider two cases. First, we  suppose  that users are requesting for different content, i.e. user 1 (user 2) requests for $\libraryf{1}$ ($\libraryf{2}$) to relay R$_{\B}$ (R$_{\W}$). Since each R has one encoded packet of the requested file in cache, S should send  one encoded packet to each R. Hence, the number of required backhaul transmissions, i.e. the realization of $l$  of the rv $     L^{\text{MDS}} $ takes value
  \begin{equation}
    l_1^{}= 2.
  \end{equation}
     Each user is able to reconstruct the file by receiving one encoded packet directly from the cache and the other forwarded by the relay. If, instead, both users request  for the same content, 
S can only  transmit the encoded packet $\encodedf{i}^{(3)}$ to both and they will successfully decode  the
requested content. In this case, the number of packets to  transmit is \[l_2  = 1. \] Combining the two cases, $\mathsf{L}$ in MDS  evaluates to
 \begin{equation}
   \begin{aligned}
       \load^{\text{MDS}}  & =  \sum_i p_L(l_i) \,   l_i= \frac{1}{2} \ l_1 + \frac{1}{2} \ l_2  = 1 + \frac{1}{2} =  \frac{3}{2}
\end{aligned}
\end{equation}
where we   sum over the $i$ possibilities on how users can request for the library content. They ask for different  files with probability
${p(l_1 ) = {1}/{2}}$ while they ask  for the same   file with probability  ${p(l_2) = {1}/{2}}.$
\end{example}

\vspace{-.7em}
\subsection*{Edge Coded Caching Scheme}
In the ECC scheme, caches are filled with non-encoded fragments, while the encoding takes place in the delivery phase. S creates coded delivery opportunities so that with a  unique transmission both relays are able to recover the desired information  also when   different content is requested.

In the placement phase, each R fills its cache with $F_j \, \nfrac $ exclusive fragments of file   $f_j$ so that relays have different fragments of the same file. S is aware of which content has been stored in each R. In the second phase, users make their requests to the corresponding R. The delivery can be split into three stages. In the first stage, a user receives fragments  directly from the cached content of the associated R. In the second stage,    \ac{S} is informed of users requests and provides missing content over the shared bachkaul link by creating coded multicast opportunities transmissions  when is possible. In this stage the relays decode the transmission and forward the desired packets to users.
In the third and last stage, S sends the remaining content in a non-encoded transmission, and relays forward this to users.

A coded multicast opportunity allows both relays to retrieve file fragments with a single transmission. In particular,  \ac{S} creates a coded packet by XORing two fragments (i.e., a bitwise operation). S picks a fragment of a file requested at {R}$_{\B}$ and present in cache of  {R}$_{\W}$ and vice-versa and combines them for delivering.  In this way,  each R receives a coded packet which is composed by a fragment present in own memory and a desired fragment. Each R by XORing the received packet with the corresponding  fragment in cache obtains a fragment of the requested file. 
ECC generates a   gain over the MDS caching scheme whenever relays have disjoint requests.   Let us clarify the last statement by considering the setting discussed in Example 1.

 \begin{example}
Let us assume to have one user per relay and the following cache placement: ${\mathcal{Z}_{\B} =\{\libraryf{1}^{(1)}, \libraryf{2}^{(1)}\} \text{ and }  \mathcal{Z}_{\W} =\{\libraryf{1}^{(2)},\libraryf{2}^{(2)}\} }$. 

Consider   first the case where   users request for different content. For example, user 1 (user 2) requests for   $f_1$ ($f_2$) to R$_{\B}$ (R$_{\W}$). During the first stage, user $i$ receives $\libraryf{i}^{(i)}$  from the cache of the related R. At the second stage, S sends  the following coded packet $p = \libraryf{2}^{(1)} \oplus \libraryf{1}^{(2)}$. Thus the number of packets transmitted over the backhaul,  $l_1$, is
\setlength{\belowdisplayshortskip}{0.2pt} 
\begin{equation}
  l_1   = 1.
\end{equation}
{R}$_{\B}$ ({R}$_{\W}$) reconstructs the   missing fragment by computing ${p \oplus \libraryf{2}^{(1)}}$${(p \oplus \libraryf{1}^{(2)})}$.
 Similarly, when users request for the same file, both are satisfied with a single coded transmission, i.e.  \[l_2 =1.\] In the ECC scheme, $\mathsf{L}$ is then
   \begin{equation}
   \begin{aligned}
     \load^{\text{ECC}} & = \sum_i p_L( l_i) \,  l_i  =  \frac{1}{2} \ l_1+ \frac{1}{2} \ l_2 = 1
   \end{aligned}
   \end{equation}
 \end{example}

With the presented examples, we observe that there exists a gain in the ECC scheme whenever there are requests for different files at relays. To understand the potential of this gain, we derive $\mathsf{L}$ in both schemes in a more general setting. To this aim, we start by recalling some useful results of the BiB problem, which will be later applied to our derivations.

\begin{figure}[!t] \centering
\includegraphics[width=0.4\textwidth]{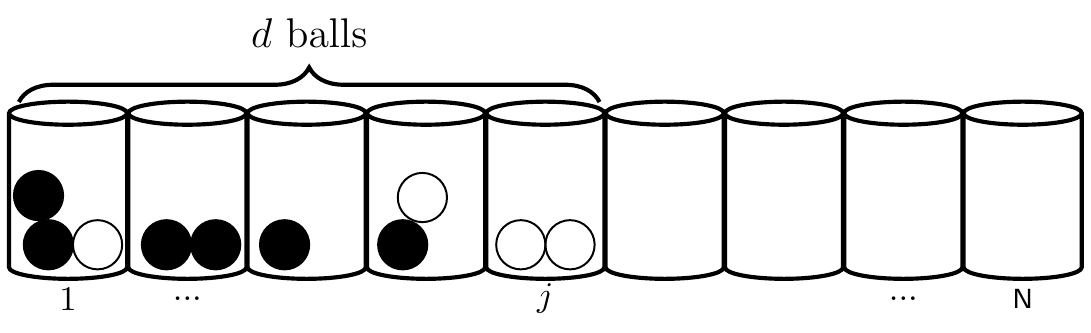}
\caption{Caching requests represented as  the  BiB problem. Bins represent files while balls represent users' requests. The occupancy problem lies on calculating the probability of having  exactly $j$ bins  not empty after throwing $d$ balls, i.e. the probability that $d$ users request  exactly for $j$ different files. }
\label{fig:ballsinbins1}
\end{figure}

\begin{figure}[!t] \centering
 \includegraphics[width=0.44\textwidth]{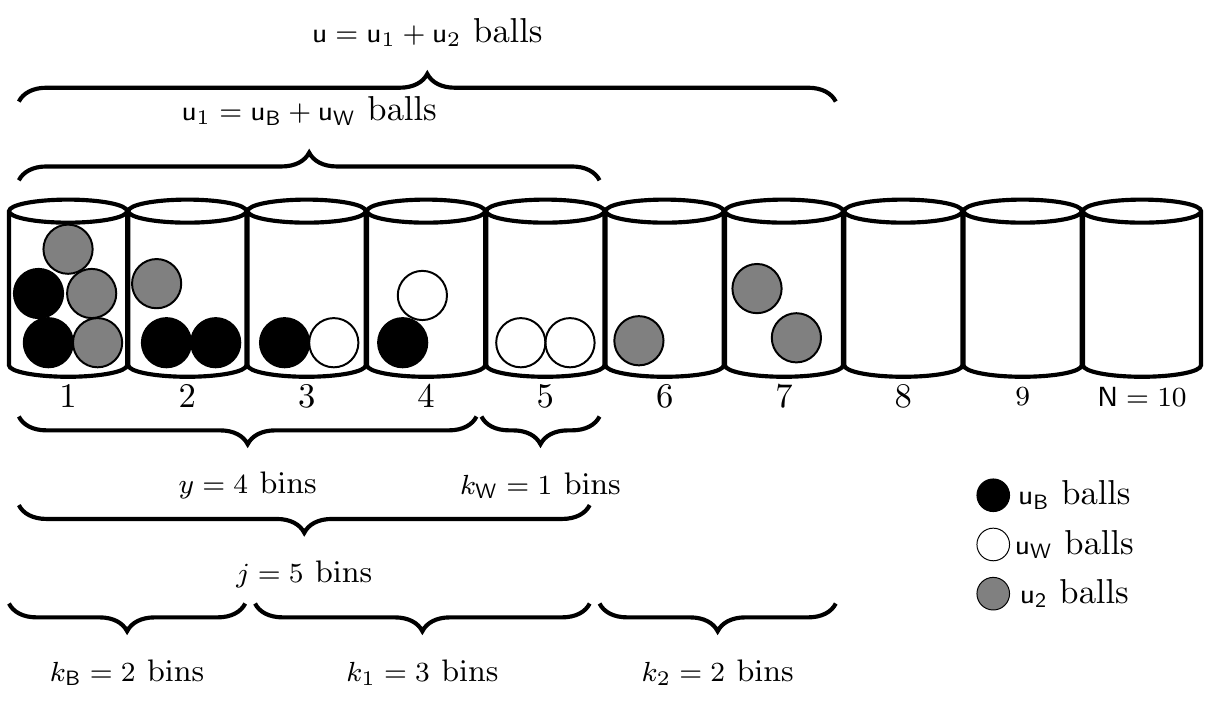}
 \centering \caption{Requests  represented as  the BiB problem.
Black balls represents requests from users connected only to R$_{\B}$, while balls request only to  R$_{\W}$ while gray balls represents request from users connected to both relays. }
 \label{fig:bib_ex}
\end{figure}

\section{Balls into bins problem applied to caching} \label{sec:BiBcaching}
To instantiate such calculations, it is convenient to map our setting onto a \acl{BiB}  setup. The general \ac{BiB} problem, see e.g. \cite{Kotz}, consists in independently throwing $d$ balls into $\N$ bins. As illustrated in  Fig.~\ref{fig:ballsinbins1}, this can be cast to our caching problem by having each bin associated to a file of the library, and by having balls which represent user requests. Following this parallel, the possibility for more balls to land into the same bin corresponds to have multiple users asking for a common library element.
 A first useful result is given by the probability of having exactly $j$ bins out of $\N$ non empty given that  $d$ balls are thrown uniformly at random, which  was derived in \cite{Recayte:outage} and can be written as
\vspace{-.6em}
 \begin{equation}\label{eq:pdelta}\setcounter{equation}{0}
     p_J(j |d)    = \frac{{\N \choose j} \,\mathit{S}(d, j)\, j! }{\N^{d}}
 \end{equation}
where $\mathit{S}(d, j)$ is the Stirling number of second kind, i.e.,  \[ \mathit{S}(d, j) = ({j!})^{-1} \sum_{i=0}^d (-1)^i \binom{d}{i}  (j-i)^d. \]

In our setup, we further  need to differentiate requests made to  {R}$_{\B}$    from those made to {R}$_{\W}$ and, similarly, requests made by users connected to one relay (set $\mathcal{U}_1$), from those made by  users connected to two relays (set $\mathcal{U}_2$).  To this aim we distinguish requests at different relays, as illustrated in Fig. \ref{fig:bib_ex}, by considering balls of  two different colors, e.g. black and white balls. A bin containing black and white balls indicates that the same file is required at both relays. Having the $i$-th bin with only black (white) balls implies that the $i$-th file was requested only at relay R$_{\B}$ (R$_{\W}$).

Following this approach, a useful result is offered by the  \emph{multivariate occupancy problem}  assuming that there are $\N$ bins and that $\us_{\B}$ black balls have been thrown and have occupied $j$ different bins. The probability that, after throwing $\us_{\W}$ white balls, there are exactly $k_{\B}$ bins containing only black balls and $k_{\W}$ bins containing  only white balls is \cite{Kotz}
\begin{equation}\label{eq:pwb}
\pwb(j, \kbinblack, \kbinwhite, \nballwhite) = {\binom{j}{\kbinblack} \binom{\N\tiny{-}j}{ \kbinwhite}} \frac{\Delta^{b_{\mathsf{W}}}  \, \mathbf{0}^{u_{\W}}}{ \N^{\nballwhite} }
\end{equation}
where  $b_{\mathsf{W}}$ is the number of bins containing the $\us_{\W}$ white balls, i.e. $b_{\mathsf{W}} =  j - \kbinblack  +\kbinwhite $
and the quantity
\[ \Delta^{m}  \, \mathbf{0}^n  := \sum_{i=0}^m (-1)^i {m\choose i} (m -i)^n \]
is known as \emph{difference of zeros} \cite{Kotz}.
Note that, in our setting,  $\pwb(j, \kbinblack, \kbinwhite, \nballwhite) $ provides the probability that exactly  $k_{\B}$ files are requested only to relay R$_{\B}$ and
$k_{\W}$ files are requested only to relay R$_{\W}$ when in total there are $\us_1= \us_{\B} +\us_{\W}$ users connected to exactly one relay.

\vspace{-.7em}
\section{Average Backhaul Transmission Load} \label{sec:Gain_edge}
 \begin{table}[!t]
     \centering
        \caption{List of some of the random variables}
     \begin{tabular}{l l l}
  \textbf{rv} & \textbf{Definition}  & \textbf{Alphabet}    \\ [.3em]
  \hline
 $J$                &  $|\demands_{{\zoneone}}|$        &  $\{1, ...,  \beta_{\Jdiff} = \min(\us_{\zoneone}, \N)\}$ \\[.3em]
 $\Ydiff$           &  $|\demands_{{\B}}|$  & $\{1, ...,  \beta_{\Ydiff} = \min(\us_{\B}, \N)\}$\\[.3em]
 $K_\mathsf{B}$     &  $|\demands_{{\B}} \backslash \demands_{{\W}}|$
           &   $\{\alfB = \max(0, y - \nballwhite + \kbinwhite ), $\\
           & & \quad  \ \quad \quad  \quad  $...,  \beta_{\mathsf{B}} = \min(\nballblack, \N) \}$ \\[.3em]
 $K_\mathsf{W}$     & $|\demands_{{\W}} \backslash \demands_{{\B}}|$
           &  $\{0, ... ,\beta_{\mathsf{W}} = \min(\nballwhite, \N - \beta_{\mathsf{B}}   ) \}$ \\[.3em]
 $K_{\zoneone}$     & $|\demands_{{1}} \backslash \demands_{{2}}|$  &  $\{ \alfGone=\max(0, j  - \us_{2}),$\\
 & & \quad \quad  \quad  \quad$ ..., \beta_{ {\zoneone}}  = \min (\us_{\zoneone}, \N) \}$\\[.3em]
 $K_{\zonetwo}$     & $|\demands_{{2}} \backslash \demands_{{1}}|$ & $\{0,...,  \beta_{{\zonetwo}} = \min{ (\us_{\zonetwo}, \N - \minGone)} \}$ \\[.3em]
 $Z$  &  $\min\{K_\mathsf{B}, K_\mathsf{W}  \}$  &  $\{ 0, ..., y \}$ \\
    \hline
     \end{tabular}
     \label{tab:rv}
\end{table}
Leaning on the parallel with the BiB problem, we now derive  the mean number of packets/fragments that S needs to send via the backhaul link  to satisfy $\us$ requests.

For convenience, we list in Table~\ref{tab:rv} the rvs  needed for our derivations together with their definition and alphabet. The first column indicates the notation of the rv, the second   its definition and the last   its alphabet. For instance, the rv $J$ denotes the number of different files requested by $\us_1$ users connected to only one relay ($\mathcal{U}_1$) while $K_{\B}$ denotes the number of different files requested exclusively at {R}$_{\B}$. Instead, the notation $\demands_{{\B}} \backslash \demands_{{\W}}$ indicates the set difference and that is the set of file requested at R$_{\B}$ but not requested at R$_{\W}$. Let us clarify all the mentioned quantities with an example.
\begin{example}
Let us refer to Fig.~\ref{fig:bib_ex} which íllustrates a library of $\N=10$  files (bins) and $\us  = 17$ users (balls). There are  $\us_{\B}=6$ users  connected only  to R$_{\B}$ (black balls), $\us_{\W}=4$ only to R$_{\W}$ (white balls), while $\us_2 = 7$ are connected to both relays (grey).

We have that  $\us_{\B}=6$ users   requested  in total $y=4$ different files  (represented by the four bins with black balls). Users in $\mathcal{U}_2$  asked in total for three files and are represented by  bins 3, 4 and 5. The  files requested by the $\us_1$ users connected to only one relay are in total $j=5$, i.e. the number of bins with black or with balls. Out of those, the files requested exclusively at R$_{\B}$ are  $k_{\B}=2$, i.e. the number of bins with black balls and without white balls, while those exclusively requested at R$_{\W}$ are  $k_{\W}=1$, i.e. the number of bins with white balls and without black balls. Since the minimum number of  mono-colour bins is one, then $z = 1$, i.e. $z = \min\{k_{\B}, k_{\W}\}$.

Users connected to both relays requested in total  for 4 further files,  represented by bins 1, 2, 6 and 7. Then   the  files requested only at one R are bins 3, 4 and 6 so in total $k_1=3$, while files requested only by users connected to both relays are bins 6 and 7 such that  $k_2=2.$
\end{example}
In the next derivations we assume that files are equiprobable, each file is split into $\nfrac = \N$ fragments and the number of files stored at each relay is $F_j = \M$ for all $j$.

%

\begin{figure*}[t!]
\small
\setcounter{equation}{9}
\begin{align}
\begin{split}
 \mathsf{\bar{L}}^{\text{MDS}}  = \sum_{j=1}^{\minJ} \frac{ {\N \choose j } S(\uzoneone, j) j! }{\N ^{\uzoneone}} \Big[j \big(1- \frac{\M}{\N}\big) + \big(1 -   \frac{2\M}{\N}\big)^+ \sum_{\kzoneone = \alfGone }^{j} {j \choose \kzoneone} \sum_{\kzonetwo =0}^{\minGtwo} {\N-j \choose \kzonetwo -1 } 
  \sum_{i=0}^{\N  - b_{2}} (-1)^i {\N - b_{2} \choose i} \Big(\frac{\N -b_{2} - i}{\N}\Big)^{\uzonetwo} \Big].
\end{split}
 \label{LMDS}
 \end{align}
\hrulefill
\end{figure*}

\subsection{MDS Average Transmision Load}
Let us recall that $J$ different files are requested by $\us_1$ users in $\mathcal{U}_1 $. 
 By the working principle of the MDS scheme, for each file requested, S has to send in the backhaul $\nfrac - \M$ packets, whereas the remaining $\M$ are already provided to the  user via the relay's cache.
 The overall number of packets that S  transmits to satisfy  $\us_1$  requests is then expressed by the rv
\begin{equation}\label{eq:load1mds}\setcounter{equation}{2}
\lmds  = (\nfrac - \M) \, J.
\end{equation}
To complete the analysis, we derive the number of packets needed to satisfy users connected to both relays (users in $\mathcal{U}_2$). Note that in this calculation it is needed to take into account only the $K_2$ \emph{aggregated} requests, i.e. the new files requested  by  $\mathcal{U}_2$ users but not requested by $\mathcal{U}_1$. In fact, whenever a file requested by   users in  $\mathcal{U}_2$ coincides  with a user request from  $\mathcal{U}_1$, both
requests are satisfied with the same backhaul transmission already accounted for by $\lmds$. Observing that each user in $\mathcal{U}_2$ receives in total $2\M$ different fragments of the respective file from relays, the number of packets that S has to send for each aggregated file is $(  \nfrac-2 \M )^+$
  where  ${(x)^+ := \max(0, x)}$. Note that whenever $\M \geq {\N} / {2}$, no transmission in needed.

Combining these remarks, the transmissions that S has to perform to satisfy the aggregated  requests can  be expressed as
\begin{equation}\label{eq:load2mds}\setcounter{equation}{3}
\lmdsdue  = (\nfrac - 2\M)^+ \, K_2.
\end{equation}
So that the average backhaul load $\mathsf{L}$ in the MDS is
\begin{equation}\label{eq:sumL}\setcounter{equation}{4}
  \begin{aligned}
     \load^{\text{MDS}} = \mathbb{E}\big[\lmds\big] + \mathbb{E}\big[\lmdsdue\big].
  \end{aligned}
\end{equation}
  Let us now calculate the two addends of equation \eqref{eq:sumL}.

The average transmission load for users in $\mathcal{U}_1$  can be computed by simply averaging over $J$ to obtain \vspace{-1em}
 \begin{equation}\label{eq:load1mds}\setcounter{equation}{5}
     \begin{aligned}
  \load_{1}^{\text{MDS}} & =  \mathbb{E}\big[  (\nfrac - \M)   \,  J \big] \\
  &  = (\nfrac - \M) \,  \sum_{j=1 }^{\beta_J } p_J(j|\us_{\zoneone} ) \cdot j \\
   & = (\N - \M) \sum_{j=1 }^{\beta_J  }\frac{{\N \choose j} \,\mathit{S}(\us_{\zoneone} , j)\, j! }{\N^{\us_{\zoneone} }}  \cdot j
   \end{aligned}
 \end{equation}
where the quantity $ p_J(j|\us_{\zoneone} )$  was derived with  BiB occupancy problem, see \eqref{eq:pdelta}.

The average transmission load given by the aggregated
 files requested by  $\mathcal{U}_2$    can be computed by conditioning to $J$, i.e.

\begin{equation}\label{eq:load2mds}\setcounter{equation}{6}
     \begin{aligned}
  \load_{2}^{\text{MDS}} & =   \mathbb{E}_J\Big[  \mathbb{E}\big[ (\nfrac - 2\M)^+ K_2 |J\big]  \Big]. \\
   \end{aligned}
 \end{equation}
Let us first focus on the inner expectation, and derive the conditional pmf $p_{K_2}(k_2|j)$, i.e. the probability that users connected to both relays request for exactly $k_2$ new files given that $j$ different files have been requested by users connected to one relay.  To help the reader, we refer to   Fig.~\ref{fig:bib_ex}
the sought probability can be computed in the BiB setup as the probability of having $j + \kbinred $ non empty bins after throwing $\us_2$ (grey) balls, conditioned on having already $j$ non empty bins occupied by $\us_1$ balls.
As discussed, this results is offered by the multivariate occupancy problem, and we have

\begin{equation}\label{eq:pkred}
  p_{K_2}( \kbinred |j ) =  \sum_{k_{\zoneone} = \alpha_1}^{j} \pwb(j, k_{\zoneone}, \kbinred, \us_{2}),
\end{equation}
where   the correspondent number of file requested at both relays is  ${b_{2} =  j - k_{\zoneone} + k_{\zonetwo}}$.  In \eqref{eq:pkred} we are summing up all the possible values that  $k_{\zoneone}$  can assume (i.e. files exclusively requested at R$_{\B}$  represented by bins with only black balls).
Accordingly,
 \begin{equation}\label{eq:load2mds}
    \begin{aligned}
&  \load_{2}^{\text{MDS}}  =   \mathbb{E}_J\Big[ (\nfrac - 2\M)^+ \sum_{k_2=0}^{\beta_2} k_2 \, p_{K_2}(k_2 |J) \Big]  \\
&  = (\N-  {2\M})^+ \sum_{j=0}^{\minJ} p_J(j|\us_{\zoneone})   \sum_{\kbinred =0}^{\minGtwo }   \kbinred \sum_{k_{\zoneone} = \alfGone }^{j} \pwb(j, k_{\zoneone}, \kbinred,  \us_{2} ).\\
 \end{aligned}
 \end{equation}

By inserting   \eqref{eq:load1mds}  and  \eqref{eq:load2mds} onto \eqref{eq:sumL}  we obtain $\load^{\text{MDS}}$ and normalizing by the number of fragments $\N$, we have that the normalized average transmission load, $\mathsf{\bar{L}}^{\text{MDS}} =  \load^{\text{MDS}} / \N $, in the MDS scheme is given in \eqref{LMDS} at the top of the page.

\vspace{-.7em}
\subsection{ECC Average Transmission Load}
\begin{figure*}[b!]
\small
\setcounter{equation}{15}
 \hrulefill
\begin{align}
\begin{split}
 &\small{ \bar{\load}^{\text{ECC}}  =  \sum_{y=1}^{\minB} \frac{{\N \choose y} \,\mathit{S}(\nballblack, y)\, y! }{\N^{\nballblack}}  \Bigg\{
   \sum_{\kbinwhite = 0}^{\minW}  \binom{\N-y}{\kbinwhite-1}   \Big[ (y + \kbinwhite ) \Big(1 - \frac{\M}{\N}\Big) \Big]  \Bigg[ \sum_{\kbinblack= \alfB }^{y} \binom{y}{\kbinblack}  \sum_{i = 0}^{\N - b_2 } (-1)^i \binom{\N - b_2 }{i} \Big( \frac{\N - b_2 - i}{\N}\Big)^{\us} \Bigg]   - \frac{\omega_1}{\N} } \\
   &  \sum_{z= 1}^y \ z \Bigg[ \binom{y}{z} \sum_{\kbinwhite=z}^{\minW -y - z} \binom{\N-y}{\kbinwhite}  \sum_{i=0}^{\N- \kzw} (-1)^i \binom{\N-\kzw}{i} \Big(\frac{\N -\kzw-i }{\N}\Big)^{\nballwhite} +  \binom{\N-y}{z}  \sum_{\kbinblack= z+1}^{\min (y, \minB)} \binom{y}{\kbinblack} \sum_{i=0}^{\N- \kzb} \binom{\N- \kzb}{i} (-1)^i   \\
    &  \Big( \frac{\N- \kzb-i }{\N}\Big)^{\nballwhite}  \Bigg]   \Bigg\}
     + \Big(1-  \frac{2\M}{\N}\Big)^+  \sum_{j=1}^{\beta_J} \frac{{\N \choose j} \,\mathit{S}(\us_{\zoneone}, j)\, j! }{\N^{\us_{\zoneone}}}     \Bigg[ \sum_{k_1 = \alpha_1}^j {j \choose k_1}  \sum_{\kzonetwo =0 }^{\minGtwo}    {\N-j \choose \kzonetwo-1} 
   \sum_{i=0}^{\N - b_{2}} (-1)^i  {\N - b_{2} \choose i} \Big(\frac{\N - b_{2}- i}{\N} \Big)^{\uzonetwo}\Bigg]  .
\end{split}
 \label{LECC}
 \end{align}
\end{figure*}

Let us start by considering users in $\mathcal{U}_1$. Since $\M$ fragments of the requested files are obtained from the relay's cache then each user needs ${\nfrac -\M}$  additional fragments.
Let us calculate the number of packets that S should transmit to satisfy these requests by considering the coded caching opportunities. Denoting by $Y$ the rv counting the number of different files requested by the $\us_{\B}$ users connected only to R$_{\B}$ and by $K_{\W}$ the rv counting the number of files exclusively requested by the $\us_{\W}$ connected only to  R$_{\W}$ and not requested to R$_{\B}$, in total users have to receive  $ (\nfrac-\M)(Y+ K_{\W})$ fragments in order that all their requests are satisfied. However, note that   transmissions given by the coded opportunities should not be counted. As for   Example 2, a coded transmission opportunity take places each time that  a file is requested at one relay and not in the other and vice-versa.
   S by  XORing the corresponding content present at each cache can  make a transmission useful to both relays. Each coded transmission opportunity allows the S to generate $\omega_1$ XORed packets involving the two files. In particular,
\begin{equation}
 \begin{aligned}
   \omega_1 & = \mathsf{ \min(M , N -M )},\\
          \end{aligned}
\end{equation}
where $\omega_1$ is the number of fragments per file   combined in a coding opportunity and it depends on the cache size. For each transmission opportunity,  when $\M \leq {\N}/{2} $; then  in total $\M$ fragments per  file  are XORed, whereas if  $\M> {\N}/{2}$ then  requests are satisfied by combining $\N-\M$ fragments per file.

In summary, each coded transmission opportunity allows S to combine $\omega_1$ packets where a packet is formed by two encoded fragments.
Accordingly, the overall number of transmission needed in the backhaul to serve users in $\mathcal{U}_1$ is
\begin{equation}\label{eq:load1ecc} \vspace{-.05em}
\leccone  = (\nfrac - \mathsf{M}) \, (Y + K_{\W}) -  \omega_1 \, Z
\end{equation}
where $Z$ is the rv denoting the number of coded opportunities.

Let us now consider the users connected to both relays, i.e. the set $\mathcal{U}_2$. In this case, we simply observe that no gain opportunity emerges from the aggregated requests by such users. In fact, users already receive content from both caches.
Therefore, the  value of the backhaul transmissions is the same as computed for the MDS scheme and we get  \[\lecctwo = \lmdsdue. \]
The average bachkaul transmission load of the ECC  is
\begin{equation}\label{eq:sumLECC}\setcounter{equation}{10}
  \begin{aligned}
     {\load}^{\text{ECC}} &  = \mathbb{E}\big[\leccone\big] + \mathbb{E}\big[\lecctwo\big].  \end{aligned}
\end{equation}
where we need to derive only  $\mathbb{E}\big[\leccone ]$.
 Conditioning on $Y$, we have
\begin{equation}\label{eq:l1ec}\begin{aligned}
   {\load}_1^{\text{ECC}}  &  = \mathbb{E}_Y\Big[\mathbb{E}\big[  (\nfrac - \M) \, (Y + K_{\W}) -  \omega_1 \, Z  | Y \big]\Big] \\
   &  = \mathbb{E}_Y\Big[\mathbb{E}\big[(\nfrac - \M) \, (Y + K_{\W})| Y \big]\Big]  -  \mathbb{E}_Y \Big[\mathbb{E}\big[ \omega_1 \, Z|Y \big]\Big].\\
\end{aligned}
\end{equation}
Let us first focus on the conditional distribution of $K_{\W}$. Given   $Y=y$ different files  requested from users in $\mathcal{U}_{\B}$,  the probability $p_{K_{\W}}(k_{\W}|y)$ of having exactly $k_{\W}$ files requested only at R$_{\W}$ can be derived from the BiB multivariate occupancy problem by considering all values that $k_{\B}$ can assume as
\begin{equation}\label{eq:pkw}
\begin{aligned}
 p_{K_{\W}}( \kbinwhite |y )    =  \sum_{\kbinblack= \alfB}^{y} \pwb(y, \kbinblack,  \kbinwhite, \nballwhite)
\end{aligned}
\end{equation}
where  $b_{\W} = y - k_{\B} + k_{\W}$.

Similarly, the probability $p_Z(z|y)$ of having $Z=z$ coded transmission opportunities conditioned on $Y=y$ files,  can be computed considering two disjoint events. The first is that $\us_{\B}$ users
ask exclusively for exactly $z$ files at {R}$_{\B}$ and $\nballwhite$ users  have ask at least $z$ exclusively files at {R}$_{\W}$. The second is the  probability that $\nballblack$ users   ask exclusively for more than $z$ files at {R}$_{\B}$ and $\nballwhite$ users    ask exactly $z$ exclusively files at {R}$_{\W}$. Thus, we can write
\begin{equation}\label{eq:pz}
  \begin{aligned}
  p_Z(z|y)  = \!\! \!  \sum_{\kbinwhite = z}^{\minW - y +z}  \pwb(y,z, \kbinwhite, \us_{\zoneone})  + \! \! \!  \sum_{\kbinblack = z+1}^{\min(y,\minB)}  \pwb(y,\kbinblack, z, \us_{\zoneone} )
  \end{aligned}
\end{equation}
where $ b_{\W} = y - z + \kbinwhite $ and $ b_{\B}  =  y -  \kbinblack  + z $. If we now plug \eqref{eq:pkw} and \eqref{eq:pz} into \eqref{eq:sumLECC} and we remove the condition on $Y$, we obtain
\setlength{\belowdisplayskip}{1.2pt}  \setlength{\abovedisplayshortskip}{1pt}
\begin{equation}\label{eq:l1ec}\begin{aligned}
   {\load}_1^{\text{ECC}}  &    =  \sum_{y=1}^{\minB} p_Y(y|\nballblack) \Big[  (\N - \M)  \Big( y +  \sum_{\kbinwhite = 0}^{ \minW }   \kbinwhite \, p_{K_{\W}}( \kbinwhite |y)    \Big)    \\ & - \min(\M, \N-\M) \sum_{z=0}^y  z \,  p_Z(z|y) \Big]. \\
\end{aligned}
\end{equation}

The final result in  \eqref{LECC} is obtained by adding the expression of $ {\load}_2^{\text{ECC}}$ to \eqref{eq:l1ec}
 which is the average transmission load of the ECC scheme normalized to the number of fragment ${\nfrac= \N}$, $ \bar{\load}^{\text{ECC}} = {\load}^{\text{ECC}} / \N  $ .

\section{Results}\label{sec:results}
\begin{figure}[!t]
\includegraphics[width=0.4\textwidth]{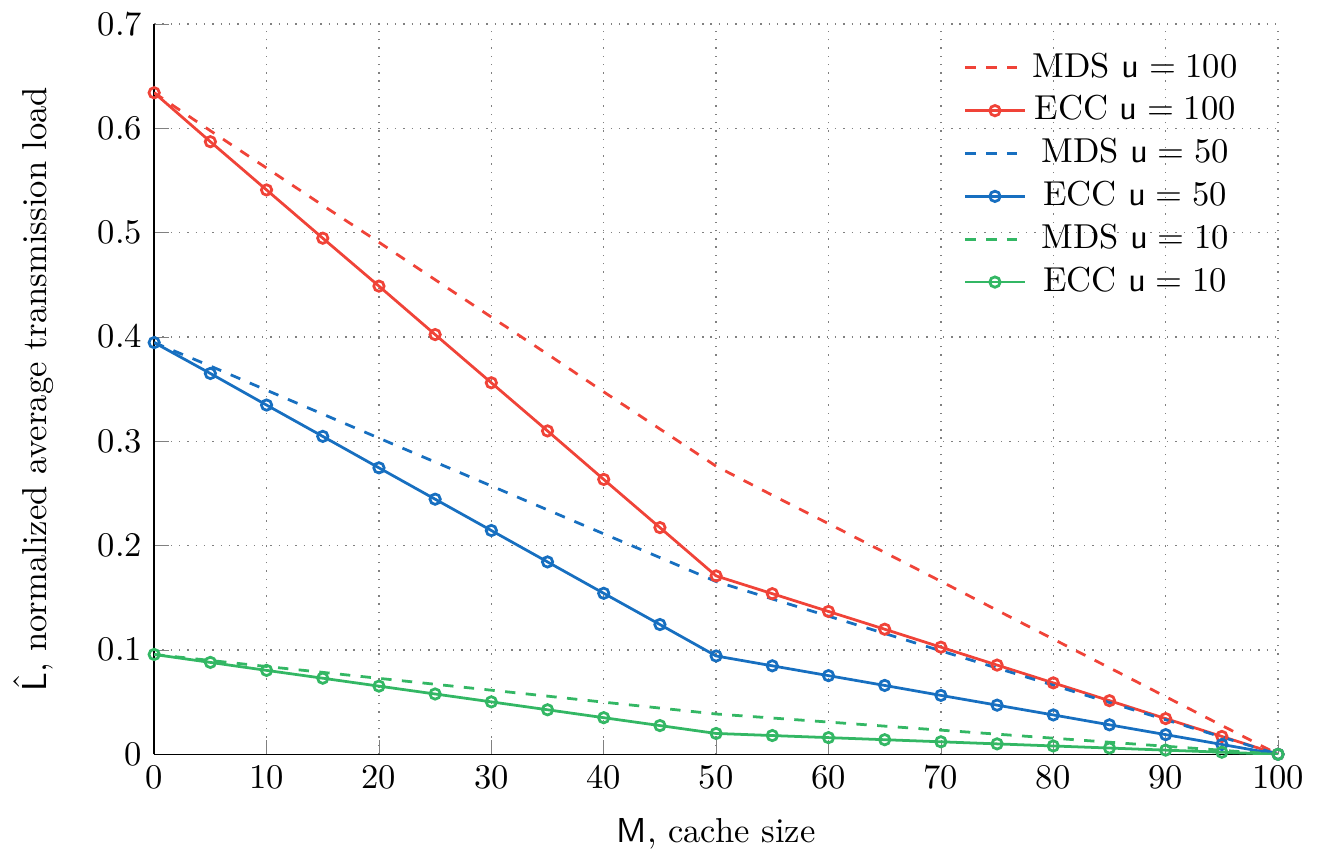}
\centering \caption{ $\hat{\load}$ versus   $\M$ for   $\us= 10, 50, 100$ for $\N= 100$.  $40\%$ of the users are connected to \ac{R}$_{\B}$,  $40\%$  to  \ac{R}$_{\W}$ and  $20\%$ to both relays. Given  $\us$, the marked and dashed curves indicate  the results   for the ECC and   the MDS scheme respectively.}
\label{fig:p_memory}
\end{figure}

We  evaluate the average backhaul load   for   the benchmark \ac{MDS} scheme and   compare it with the one of the proposed ECC scheme.
In both cases quantities are   normalized to the library size $\N$, i.e.
 ${\hat{\load} ^{\text{MDS}}  =  {{\load}^{\text{MDS}}} / {\N} } \text{ and } { \hat{\load} ^{\text{ECC}}  =  {{\load}^{\text{ECC}}}/ {\N}.} $ We assume the library size ${\N=100}$ and  20\% of the users to be connected to both relays while 80\%   to a single relay.
For simplicity, we consider  that  of half users  in $\mathcal{U}_1$ are connected only to \ac{R}$_{\B}$ and half   only to  \ac{R}$_{\W}$.

In Fig.~\ref{fig:p_memory},  the normalized average backhaul load as a function of the cache size  $\M$ for different number of users $\us$ is plotted. As shown, the ECC scheme outperforms the benchmark  MDS caching scheme for every number of users $\us$ considered.   As expected, by fixing  $\us$ requests, $\mathsf{\bar{L}}$ decreases by increasing  $\M$, since more content directly from the cache can be provided.  Given  $\M$, the gain between ECC and the MDS scheme is higher when the number of total users $\us$ is greater because more transmission opportunities take place.
The maximum gain is obtained when $\M = \N/2$, in fact,   this cache operating point encodes half of file content (the maximum portion of a file that can be combined) in a transmission opportunity.

\begin{figure}[t]
 \includegraphics[width=0.4\textwidth]{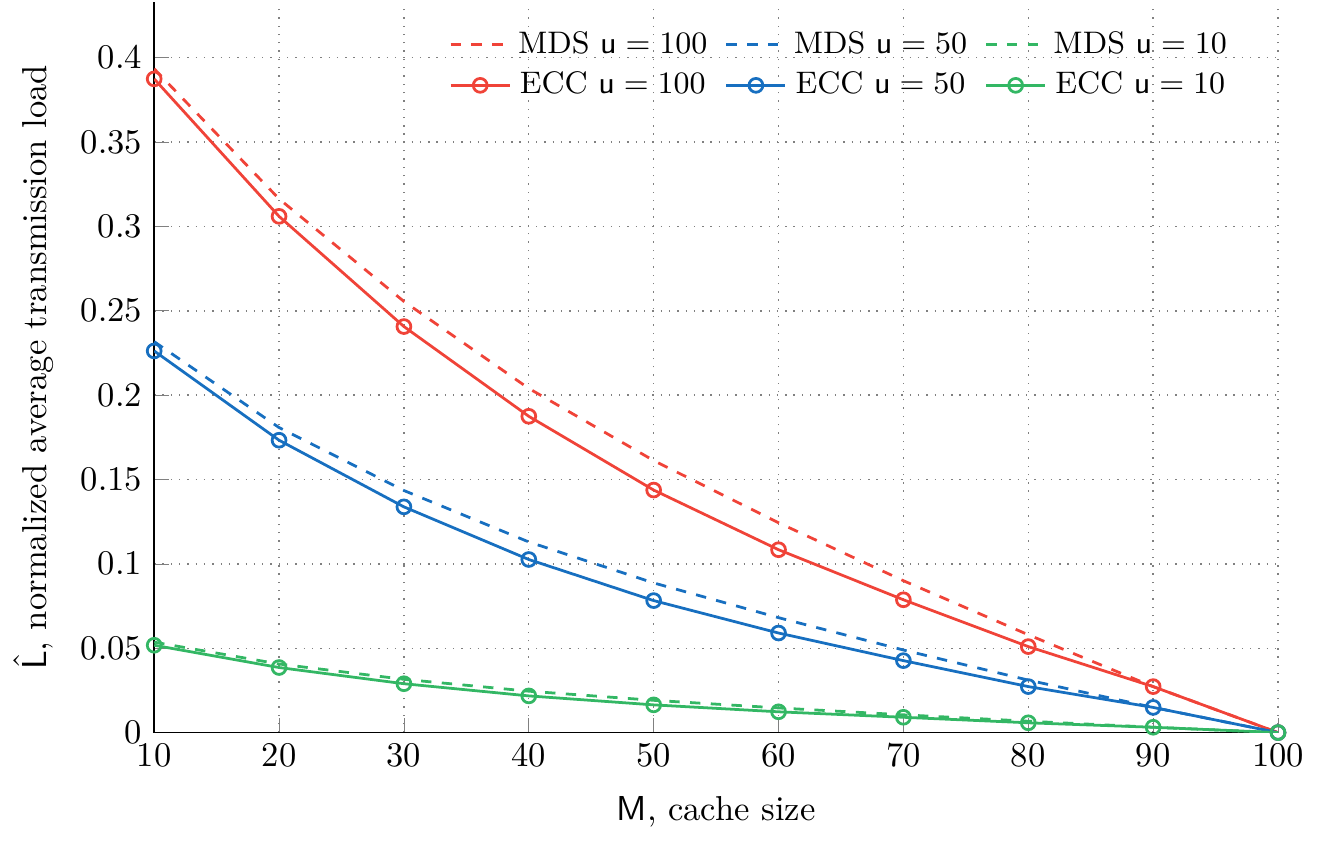}
 \centering \caption{  $\hat{\load}$ versus  $\M$  for  $\us = 10, 50, 100$ for $\N= 100$ when file requests follows a Zipf distribution with   $\alpha = 0.8$.  $40\%$ of   users are connected to \ac{R}$_1$,  $40\%$  to  \ac{R}$_2$ and  $20\%$ to both Rs. The dot marked   and   dashed curves indicate  the results obtained for the ECC and MDS scheme respectively.}
 \label{fig:load_zipf}
\end{figure}

 Motivated by the good performance obtained, we also  show  Monte-Carlo  results  when   file  request distribution is not equiprobable.
 The normalized average backhaul load in this case is reported in Fig.~\ref{fig:load_zipf}.  It is assumed that users request for content according the Zipf distribution with $\alpha = 0.80$ and  each relay optimizes own cache content according the algorithm given in \cite{bioglio:Globcom2015}.
In this set up, we can appreciate the efficiency of the caching placement due to the not uniform demands.  In fact, given the number of users $\us$, a cache size $\M$ and a scheme then $\load$ is lower than in our previous scenario. A gain on the ECC with respect to MDS is still present. Due to the lower number of coding  opportunities and due to the placement considered such gain is smaller with respect to our previous results.

%
\section{Conclusions}\label{sec:Conclusions}

We applied the Maddah-Ali caching scheme at the edge of  a two-tier heterogeneous satellite network with multiple users.  A closed-form expression of the average backhaul transmission load     for the ECC scheme was derived. The performance of the scheme was compared with those of the benchmark given by the MDS one. We quantified the nature of the transmission gain obtained by casting out problem with known results obtained in the BiB setting. Results shows a gain in terms of backhaul transmissions for any number of users considered in the system.

The relevant reduction of load backhaul transmission obtained validates  the coded caching scheme in satellite networks and  suggest its investigation in more  sophisticated  scenarios.

\section{Acknoledgment}\label{sec:ack}
This work was supported by the Federal Ministry of Education and Research (BMBF, Germany) as part of the 6G Research and Innovation Cluster 6G-RIC under Grant 16KISK020K.

\bibliographystyle{IEEEtran}
\bibliography{IEEEabrv,references}

  \flushend
\end{document}